\documentclass{iauc}
\usepackage{graphics,psfig}
\title[Wide field imaging of distant clusters]
      {Wide field imaging of distant clusters}

\author[T. Treu]{Tommaso Treu} 

\affiliation{Division of Astronomy \& Astrophysics University of
California, Los Angeles, CA 90095-1562, USA email: ttreu@astro.ucla.edu}

\pubyear{2004}
\volume{195}
\pagerange{1--8}
\date{?? and in revised form ??}
\jname{Outskirts of Galaxy Clusters: Intense Life in the Suburbs}
\editors{A. Diaferio, ed.}
\begin{document}

\maketitle

\begin{abstract}
Wide field imaging is key to understanding the build-up of distant
clusters and their galaxy population. By focusing on the so far
unexplored outskirts of clusters, where infalling galaxies first hit
the cluster potential and the hot intracluster medium, we can help
separate cosmological field galaxy evolution from that driven by
environment.  I present a selection of recent advancements in this
area, with particular emphasis on Hubble Space Telescope wide field
imaging, for its superior capability to deliver galaxy morphologies
and precise shear maps of distant clusters.
\end{abstract}

\firstsection 
              
\section{Introduction}

In the current standard cosmological paradigm (see, e.g., talks by
Benson, Moore, Springel, Tormen), clusters of galaxies form in
correspondence with the first overdensities in the early universe and
grow by mergers and by accreting material from the surrounding regions
(which I will generally refer to as the {\it field}).

In this scenario, the infall regions of clusters are believed to be
very important -- as testified by the enthusiastic participants of
this meeting. As galaxies, gas, and dark matter are accreted onto the
cluster, they are subject to interactions with the steep gravitational
potential, along with the galaxies and hot plasma already present in
the cluster. Under appropriate conditions, the outcome of these
interactions can be quite dramatic, e.g., altering the morphology and
star formation properties of infalling galaxies (e.g. Mastropietro's
talk).

In this contribution I will focus on two particular aspects of the
physics of the infall regions: i) the environmental effects on the
morphological properties of galaxies and ii) the distribution of dark
matter. As suggested by Antonaldo and the Scientific Organizing
Committee, I will focus on wide field ($>1$ Mpc radius) imaging of
distant clusters as a diagnostic tool to study these processes (for
other diagnostic tools see talks by, e.g., Balogh, Dressler,
Ellingson, Moran, Poggianti). Studying distant clusters ($z>0.1$)
provides two important pieces of information. One the one hand,
distant clusters are efficient gravitational lenses and therefore weak
and strong lensing can be used to characterize their mass distribution
(see also Schneider's talk). On the other hand, with current
technology it is possible to reach look-back times of several Gyrs,
i.e comparable to the evolutionary time-scales for these
systems. Therefore, by comparing local clusters to their
``progenitors''\footnote{With all due cautionary notes on this crucial
step.}, we can aim at empirically mapping the cosmic evolution of the
infall regions. Due to space limitations, I will restrict my
contribution to a personal selection of observational results. Most of
them are obtained with the {\it Hubble Space Telescope}, for its
ability to determine galaxy morphology and measure lensing signals.

The Hubble constant, the matter density, and the cosmological constant
are assumed to be H$_0=65$~km\,s$^{-1}$\,Mpc$^{-1}$, $\Omega_{\rm
m}=0.3$, and $\Omega_{\Lambda}=0.7$, throughout this paper.

\section{Environmental processes and physical scales}

It is empirically known that the mix of morphological types is a
function of local galaxy density (Dressler 1980) and redshift
(Dressler et al. 1997; Treu et al. 2003, hereafter T03; Smith et al.\
2004). This is generally interpreted as the combination of two
mechanisms. Firstly, high density regions are very special places in
the Universe and therefore galaxies observed there might have formed
under special initial conditions. Secondly, interactions between
galaxies, dark matter and the intra--cluster medium (i.e.\
environmental processes) are likely to transform in--falling field
galaxies from gas--rich spirals to gas--poor lenticular galaxies. In
some sense, understanding the balance between the two processes can be
thought of as a nature vs. nurture problem, although certainly the
distinction is somewhat arbitrary (the environment a galaxy is formed
in is, in a broad sense, part of the initial conditions). Therefore, I
will use the following simplified formulation of the problem. Let us
consider a general population of {\it field} galaxies and let them
fall into a cluster. When (and if) they arrive at the cluster center,
will they have changed their distribution of morphologies, as a result
of interactions, so as to match that of high density regions? If the
answer is no, then we have to invoke different initial conditions
(e.g. mass function). If the answer is yes, then we would like to know
exactly which physical mechanism is doing what.

\begin{figure}
    \centering
  \begin{minipage}[t]{0.5\linewidth} 
    \resizebox{1.0\linewidth}{3in}{\includegraphics{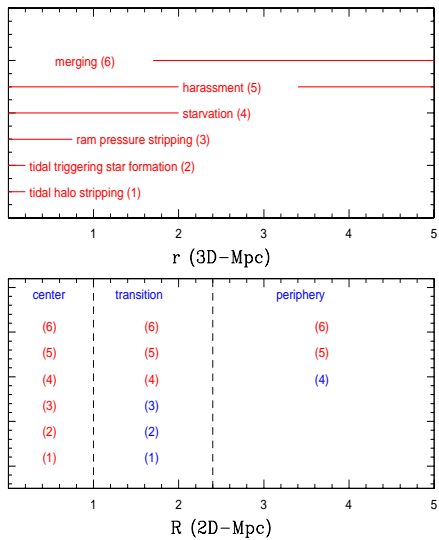}}
  \end{minipage}
\caption{Summary of spheres of influence for key physical mechanisms
in the galaxy cluster CL0024 ($z=0.39$). Top panel: the horizontal
lines indicate the radial region where the mechanisms are most
effective (in 3-D space; note that the harassment is effective in the
entire range).  Lower panel: for each projected annulus we identify
the mechanisms that can have affected the galaxy in the region
(red). The blue numbers indicate processes that are marginally at
work. See T03 for additional discussion (reproduced with permission).\label{fig:mech}} 
\end{figure}

Assuming the answer to the former question is ``yes'' or ``in part'',
in order to address the latter issue (what physical mechanism is doing
what) it is useful to rank the proposed mechanisms by estimating the
relevant time and length scales. In the upper panel of Figure~1, I show
spheres of influence for various physical mechanisms (computed as an
example for CL0024 at $z=0.39$, see T03 for definitions and
details). The lower panel shows two--dimensional spheres of influence
once projection effects, duration of phenomena, and travel times are
taken into account (see discussion of the ``back-splash''-effect in
Ellingson's and Moore's talks). It is clear that the center of the
cluster is a very busy environment, where many processes are
simultaneously at work. By contrast, the outer regions -- the {\it
transition} region around the virial radius and the {\it periphery}
beyond that -- provide a much cleaner environment, and therefore the
opportunity to separate the effects of different mechanisms.

\section{Wide field imaging}

\subsection{Ground based studies}

One of the key difficulties in interpreting wide field imaging data in
terms of galaxy population is identifying cluster members at large
radii, where the contrast over the background/foreground is
low. Abraham et al. (1996) made an important step forward in this
respect by obtaining unprecedented spectroscopic follow-up --
including hundreds of members -- of the $z=0.228$ cluster A2390. They
obtained the first observational constraints on the infall scenario,
finding that the properties of galaxies -- such as colors and spectral
features -- changed smoothly with radius, consistent with an idea of
gradual infall without significant mixing (see also Morris et al.\
1998). The recent development of wide field imagers and spectrographs
on big telescopes -- with the ability to obtain hundreds of redshift
at a time -- seems to suggest that we will see more studies of this
kind in the next years, with even larger redshift catalogs (Czoske at
this meeting presented an ongoing project collecting {\it thousands}
of redshifts per cluster).

To tackle the same problem, Kodama et al. (2001) followed a different
approach. They used photometric redshifts based on high quality Subaru
images to identify galaxies in a relatively thin redshift slice around
the cluster A851 ($z=0.39$). Although some residual contamination is
present, the use of photometric redshifts allowed them to study an
even larger sample of galaxies to fainter intrinsic
luminosities. Based on this technique, they were able to convincingly
show the filamentary structure of the accreting material. They also
detected a rapid transition in galaxy colors at a characteristic
density, possibly the signature of environmental processes influencing
infalling galaxies (see also Gray's talk).

\subsection{HST-WFPC2 studies}

The Wide Field and Planetary Camera 2 (WFPC2) revolutionized the
subject, by delivering images of galaxies in\footnote{Or behind, for
weak lensing purposes.} distant clusters with sufficient resolution to
accurately measure their shapes and morphologies (e.g. Couch et
al. 1994; Dressler et al.\ 1994; Ellis et al.\ 1997). Unfortunately,
the 4 800x800 pixel${^2}$ CCDs provide a field of view of only
$\sim2.5$ arcmin on a side, corresponding to less than a Mpc at 
$z\sim0.5$. Hence, performing wide field studies of distant clusters
requires patient and expensive mosaicing. Using contiguous pointings,
Couch et al. (1998) and van Dokkum et al.\ (1998) constructed the
morphology-density relation and color-magnitude relation reaching the
{\it transition} region of two clusters at $z\sim0.3$. They found that
the morphology density relations starts to flatten out beyond $\sim
1$Mpc and that the color-magnitude relation of lenticular galaxies is
less tight than in the cluster core (but see Andreon 1998 for an
explanation of this latter finding in terms of morphological
classification uncertainties). WFPC2 mosaics of comparable size have
been obtained now for a few clusters (e.g. talks by Dressler and
Tran).

In order to reach significantly beyond 1 Mpc in a finite number of HST
orbits, T03 adopted a sparse sampling strategy. With 39 WFPC2
pointings they covered a circle 10 Mpc in diameter, with fractional
area coverage 20-40\% in the {\it transition} and {\it periphery}
regions of cluster CL0024 at $z\sim0.39$. More than 1300 redshifts,
including those of 472 members, are available to identify cluster
members at the periphery of the cluster.

\begin{figure}
  \begin{minipage}[t]{0.5\linewidth}
    \resizebox{1\linewidth}{!}{\includegraphics{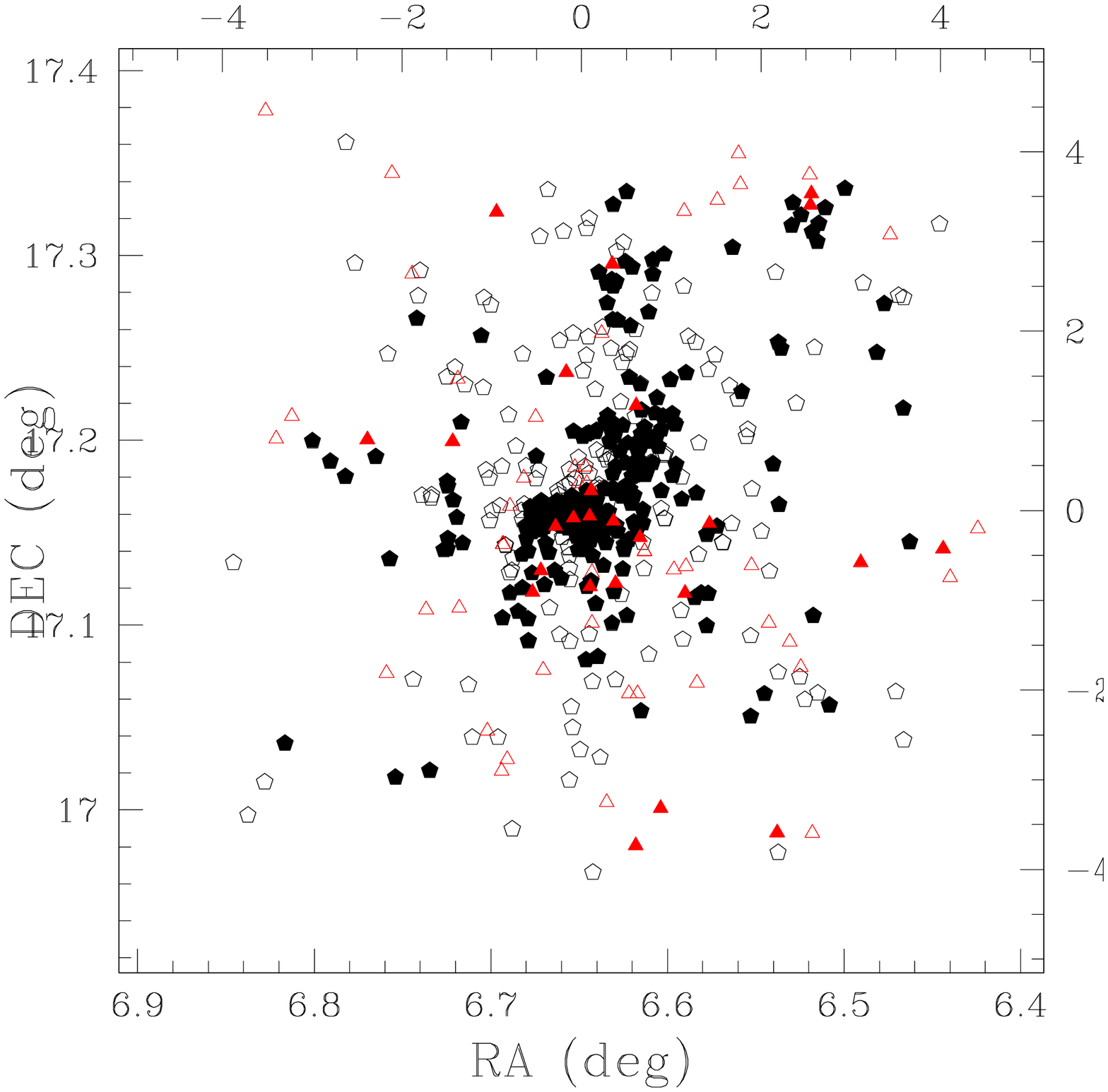}}
  \end{minipage}
  \begin{minipage}[t]{0.5\linewidth}
    \resizebox{1.05\linewidth}{!}{\includegraphics{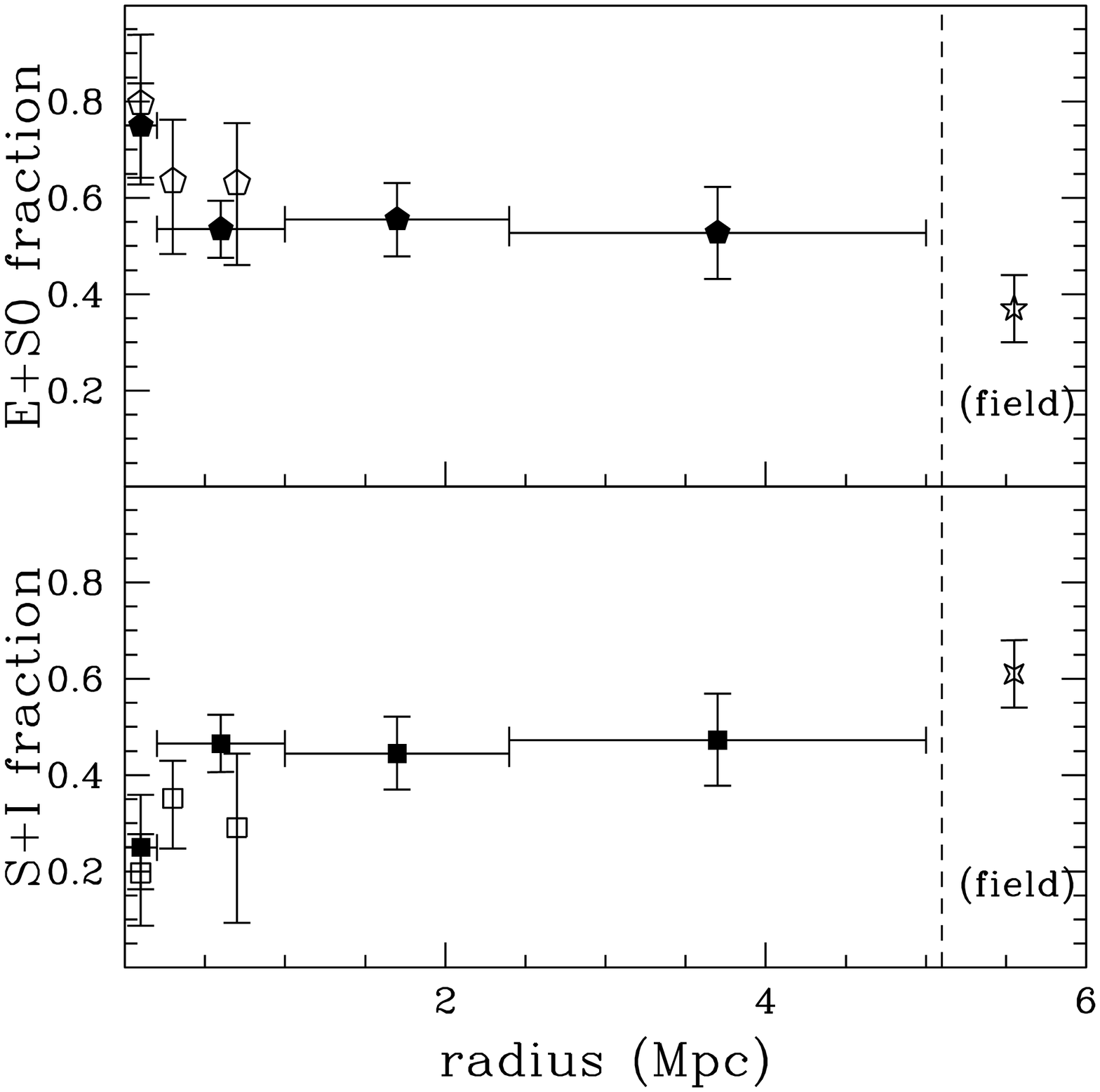}}
  \end{minipage}
\caption{Left: Projected distribution of CL0024 member galaxies. Solid
symbols indicate objects with available HST-WFPC2 images. The top and
right scale is in Mpc. Right: Fraction of morphological types
($I<21.1$) in CL0024+16 as a function of cluster radius: (upper panel)
fraction of E+S0 galaxies; (lower panels) fraction of
spirals. Fractions determined from the entire WFPC2 sample -- removing
the background statistically -- are shown as large empty points, while
fractions determined from the spectroscopic WFPC2-z catalog are shown
as solid points. Points beyond the dashed line at 5.1 Mpc are field
fractions. From T03 (reproduced with permission).
\label{fig:up0024}} \centering
\end{figure}

The left panel of Figure~2 shows an updated map of the cluster
(superceding that in T03) members. By combining HST data and redshift,
the morphology-radius relation can be measured for a single system out
to almost 5 Mpc (Figure~2, right panel; from T03, updated to include
the complete redshift catalog). Beyond 1 Mpc the fraction of
early-type galaxies declines gently with radius toward the lower
field value.  This gradient cannot be caused by mechanisms such as
ram-pressure stripping or tidal interaction with the main cluster
potential.

\begin{figure}
  \begin{minipage}[t]{0.5\linewidth}
    \resizebox{1\linewidth}{!}{\includegraphics{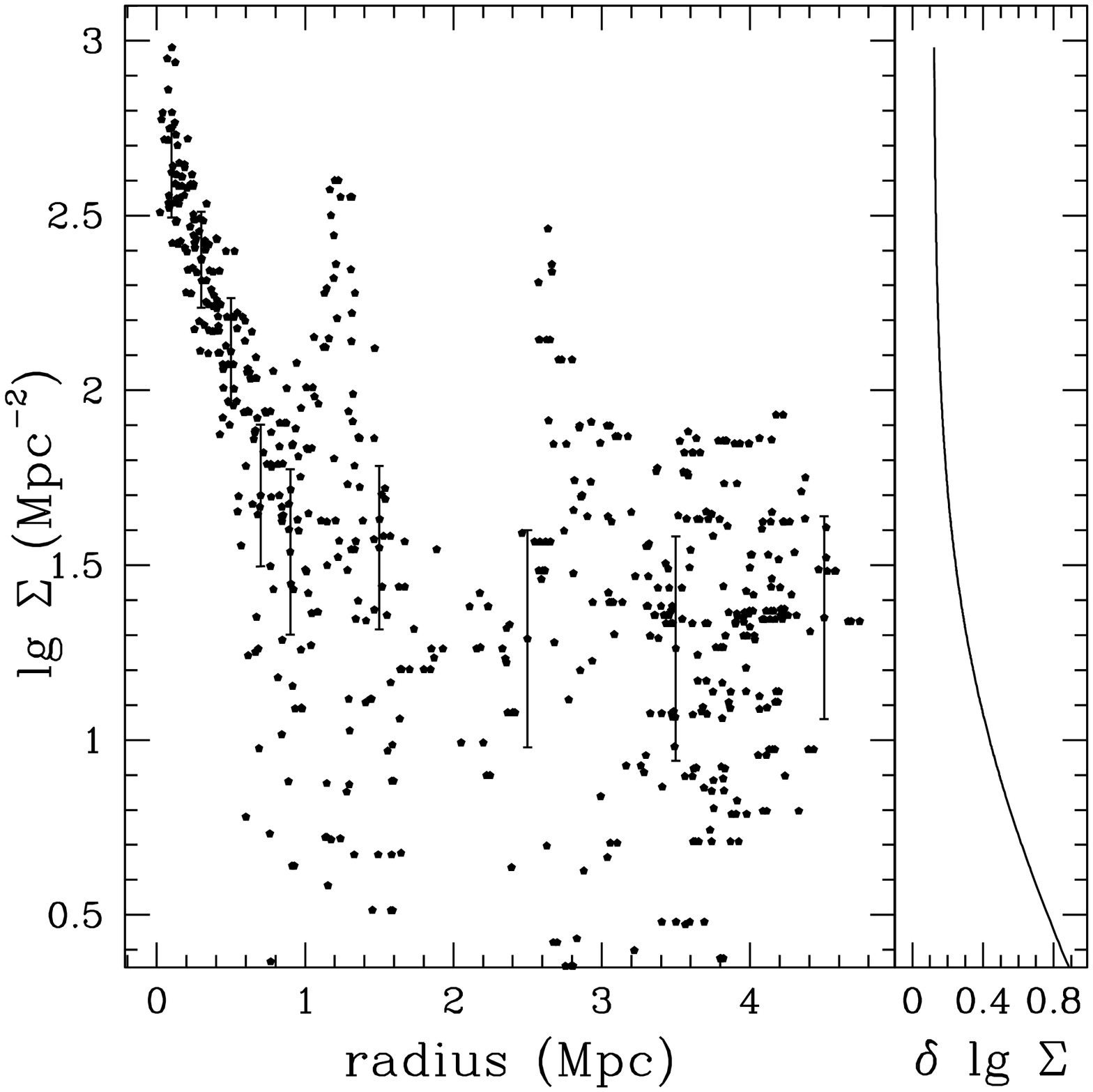}}
  \end{minipage}
  \begin{minipage}[t]{0.5\linewidth}
    \resizebox{1\linewidth}{!}{\includegraphics{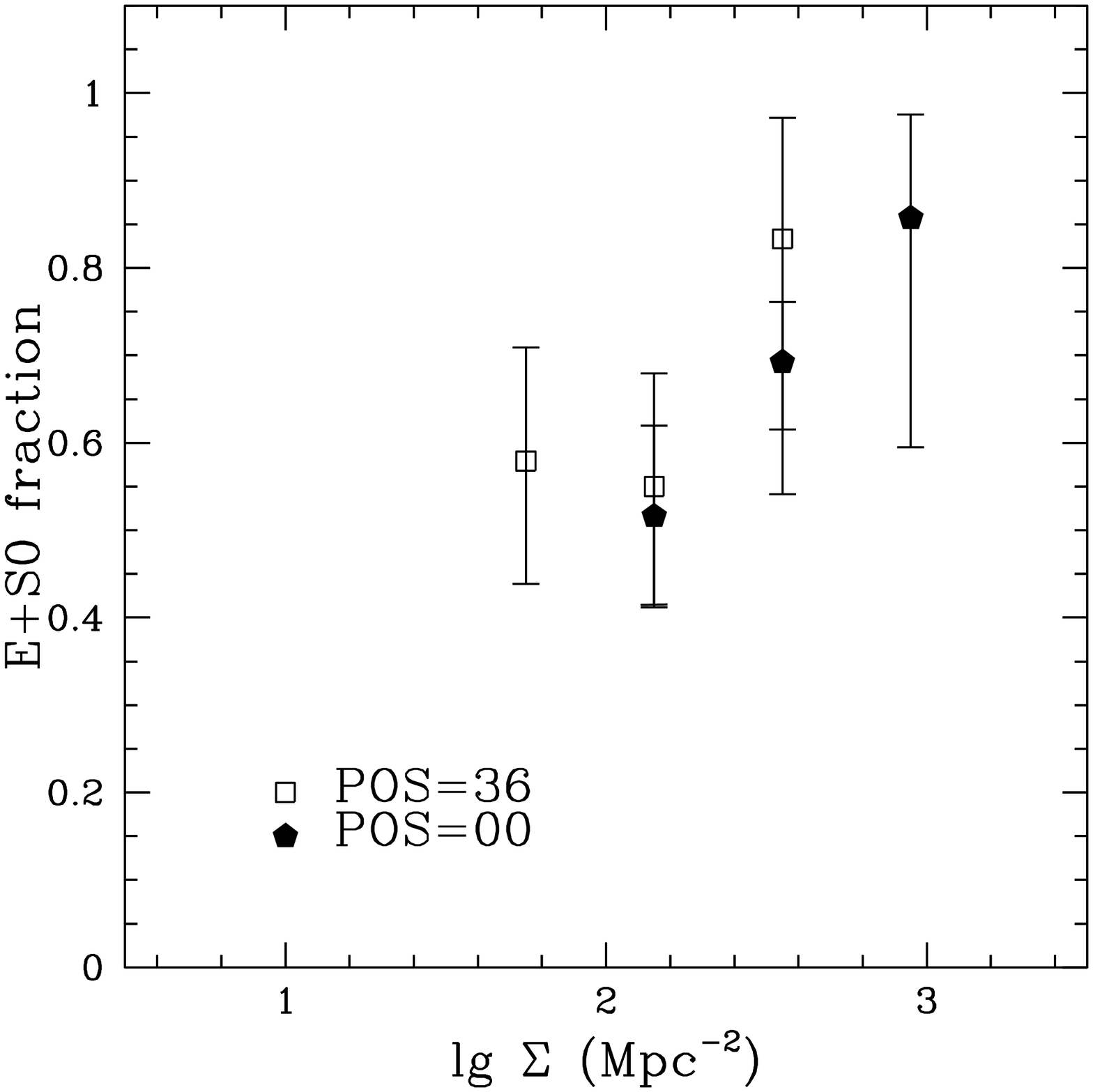}}
  \end{minipage}
\caption{Left: Local surface density vs cluster radius in CL0024
revealing the large range at a given radius and the spike in local
density at $\sim 1$ Mpc. The points with error bars indicate the
average $\Sigma$ within radial bins, and the scatter expected from
measurement errors. The expected uncertainty as a function of $\Sigma$
is shown in the right panel. Right: Fraction of E+S0 galaxies in
CL0024+16 as a function of local projected density for galaxies in the
central (solid pentagons) and in the NW overdensity region (open
squares; POS36) corresponding to the spike in local density at $\sim
1$ Mpc in the left panel.  Note the close match in the regions of
overlapping density.\label{fig:TS} From T03 (reproduced with permission).} \centering
\end{figure}

The distribution of galaxies is clearly not circularly symmetric, as
show in the left panels of Figures~2 and 3. Therefore, we can learn
more about the relationship between morphology and environmental
effects by studying the morphology-density relation as a function of
location inside the cluster (right panel of Figure~3). Remarkably, the
morphology density relation does not depend -- within the
uncertainties -- on the location inside the cluster. At a given
density, the morphological mix appears to be independent of the
cluster radius. A direct implication of this finding is that galaxy
concentrations outside the central peak are physically associated and
therefore galaxies are not accreted as individuals but rather as
groups, with their own hierarchy of morphologies (and dark matter
halos as I discuss later). At least part of the morphological
transformations could happen at the group level where galaxies are
``pre-processed'' before entering the cluster. At the group level,
different physical mechanisms are at work (e.g. merging is more
frequent, ram pressure stripping is less effective), and therefore it
is possible that different mechanisms dominate at subsequent stages.

\begin{figure}
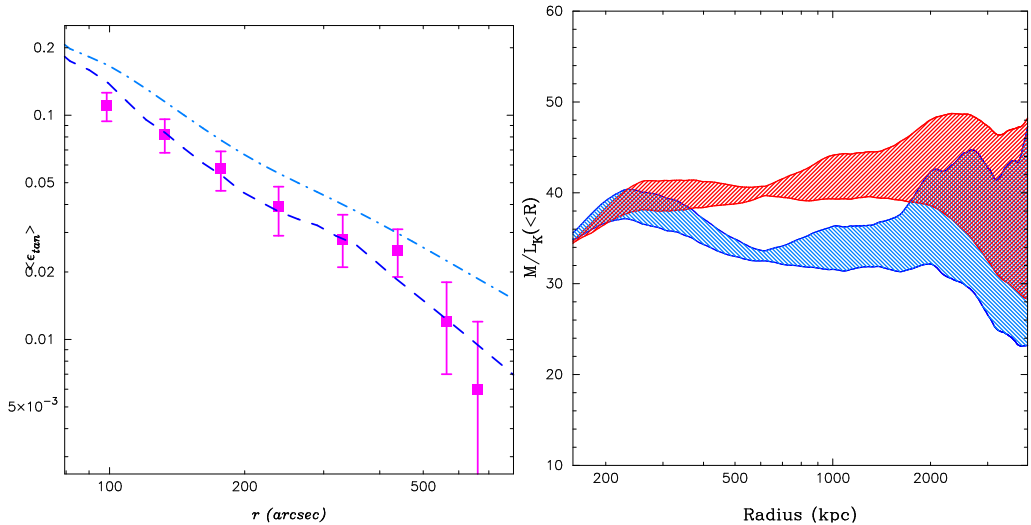

  \begin{minipage}[t]{0.5\linewidth}
    \resizebox{1\linewidth}{!}{\includegraphics{treu_f4a.eps}}
  \end{minipage}
  \begin{minipage}[t]{0.5\linewidth}
    \resizebox{1\linewidth}{!}{\includegraphics{treu_f4b.eps}}
  \end{minipage}
\caption{Left: Reduced tangential shear profile of CL0024 for the
combined WFPC2 and STIS data (magenta points with error bars).  The
dashed line is the visual representation of the reduced tangential
shear of the 2 clump NFW model that best fits both the strong and weak
lensing constraints.  The dot-dashed line corresponds to the reduced
tangential shear of the 2 clump SIS model that best fits the strong
lensing constraints but fails to fit the weak lensing
measurements. Right: $M/L_K$ ratio (rest frame solar units) of CL0024
. The (blue) lower hatched region corresponds to the $M/L$ derived for
the enclosed field-subtracted $K$-band sample and the (red) upper
hatched region that for the color-selected sample. From
K03 (reproduced with permission). \label{fig:mass}} \centering
\end{figure}

The sparse sampled mosaic of CL0024 can also be used to study the dark
matter distribution in the infall regions. A joint weak and strong
lensing analysis (Kneib et al. 2003; hereafter K03) reveals the
presence of a secondary clump of mass, coincident with the overdensity
of galaxies 1~Mpc NW of the cluster core identified in Figures~2 and~3,
with mass-to-light ratio consistent with that of the main clump.
After appropriate azimuthal averaging, the lensing analysis can be
extended to the periphery, resulting in the mass shear profile shown
in the left panel of Figure 4. The joint weak and strong lensing
constraints are consistent with a Navarro, Frenk \& White (1997) mass
profile outside the central region, and inconsistent with a singular
isothermal sphere (c.f. talks by Biviano and Mamon). The right panel
of Figure~4 shows the rest frame K-band mass-to-light ratio of the
cluster, which is approximately constant all the way out to the
periphery, indicating that the infalling material has approximately
the same fraction of mass in stars as the central regions. A similar
result for local clusters -- based on a completely independent
``caustic'' technique (Diaferio \& Geller 1997; Diaferio 1999) -- was shown as this
meeting by Rines. Remarkably, the $M/L_K$ of CL0024 is consistent with
that of local clusters, once a correction corresponding to passive
evolution of an old stellar population is applied.

\begin{figure}
  \begin{minipage}[t]{0.5\linewidth}
    \resizebox{1\linewidth}{!}{\includegraphics{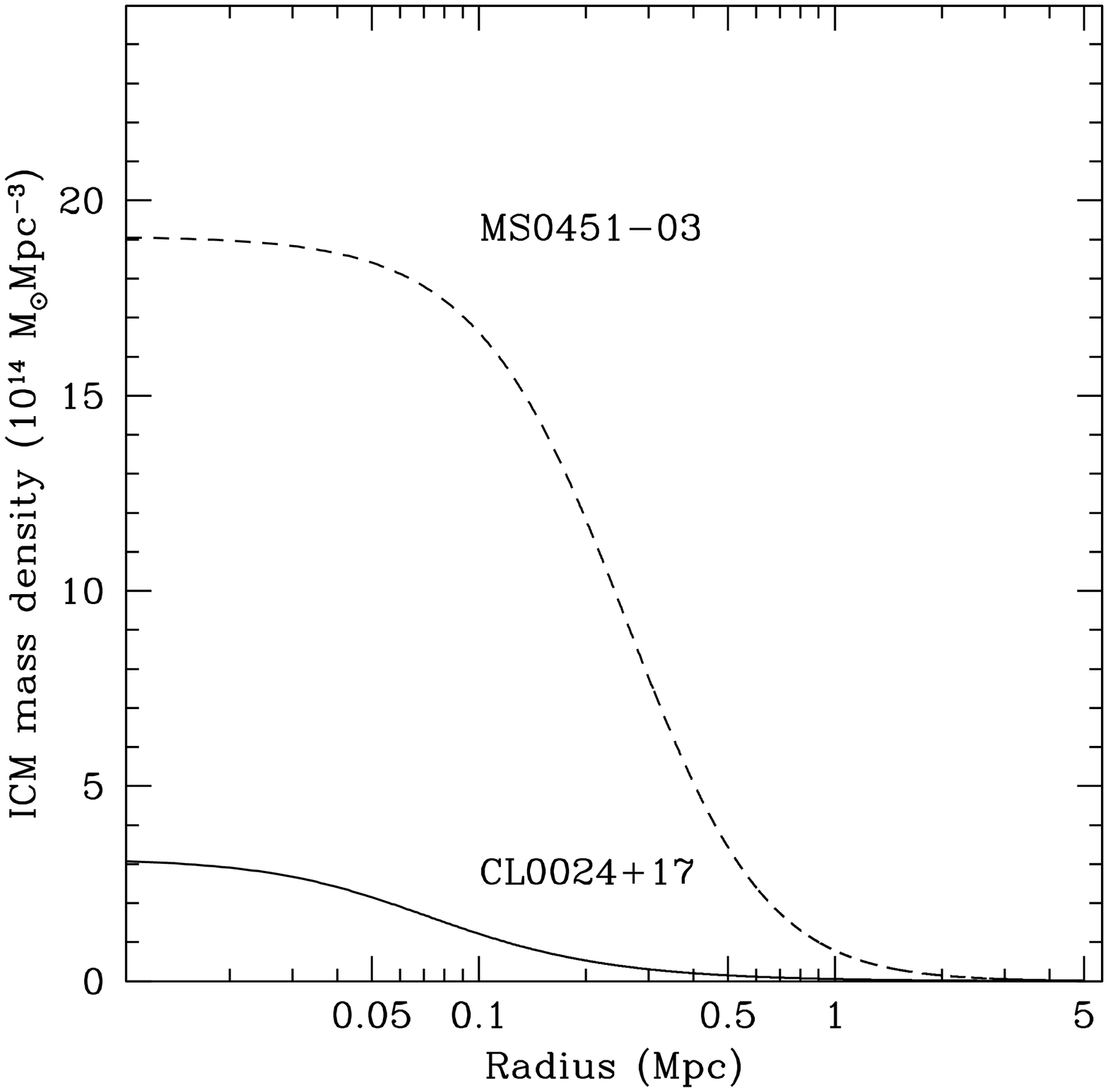}}
  \end{minipage}
  \begin{minipage}[t]{0.5\linewidth}
    \resizebox{1\linewidth}{!}{\includegraphics{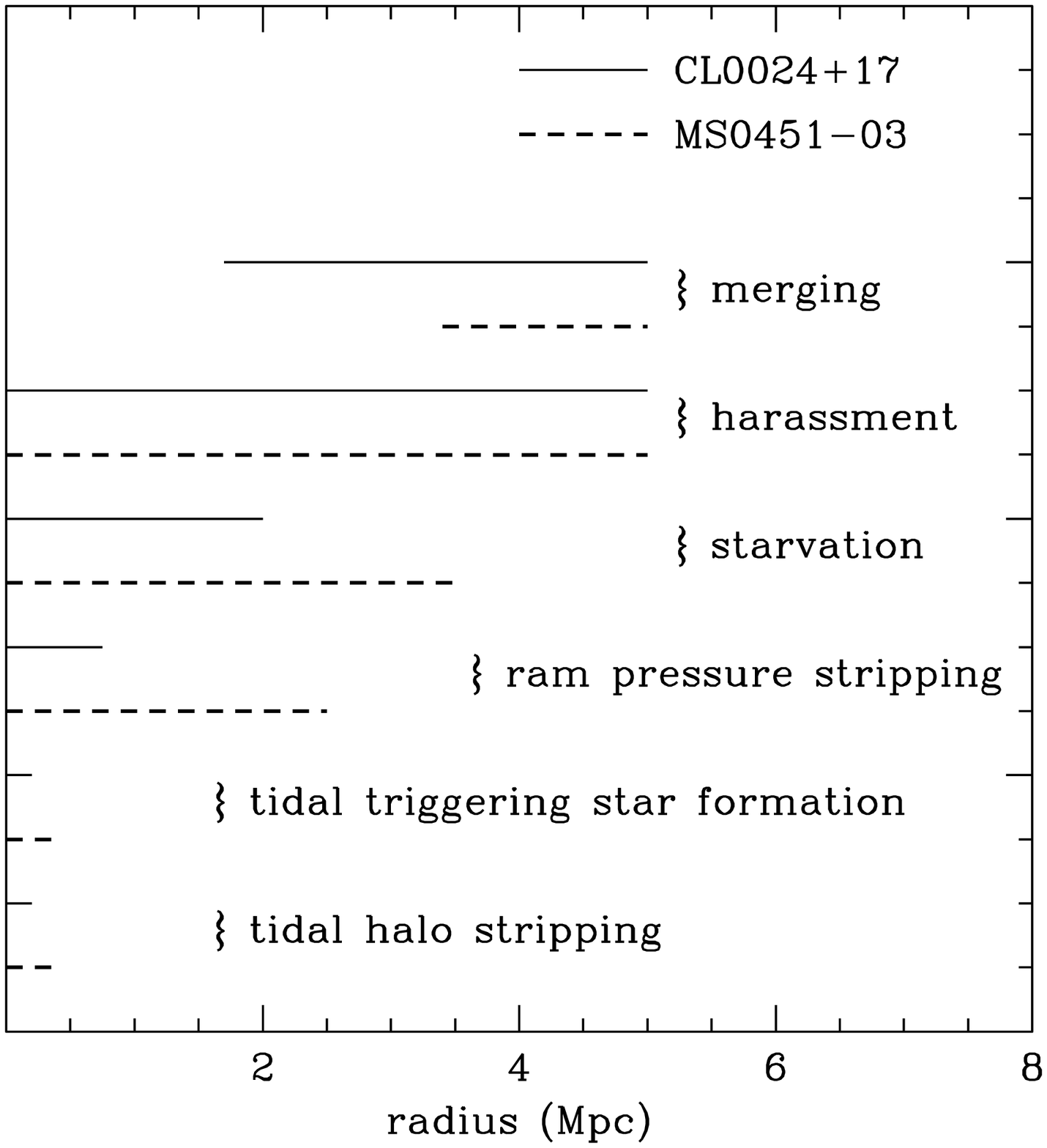}}
  \end{minipage}
\caption{Left: comparison of the mass density profile of the intracluster 
medium (ICM) in CL0024+17 and MS0451-03. Right: Comparison of
the regions of influence of key physical mechanisms for CL0024+17 and
MS0451-03. Note the dramatic enhancement in the size of the sphere of
influence of the ICM -- ram pressure stripping and starvation -- in
MS0451-03.
\label{fig:cfrclusters}} \centering
\end{figure}

\subsection{HST-ACS studies}

The successful installation of the Advanced Camera for Surveys (ACS)
on board HST has improved its ability to carry out wide field imaging
of distant clusters by a factor of 10 (considering area and
depth). Several wide field studies are currently ongoing, including
for example the ACS GTO cluster program, and the follow-up of MS1054
($z=0.83$) by Kelson et al. This improvement makes it possible to
obtain fully sampled mosaic of distant clusters over diameters
corresponding to several virial radii. It is of particular interest to
study and contrast the properties of galaxies in clusters with
different intracluster medium (ICM) properties. For example, MS0451
($z=0.54$; covered by GO-9836, PI Ellis) is a rich cluster, much more
luminous in the X-ray than CL0024. Figure~5 illustrates the different
ICM content of the two clusters and compares the spheres of influences
of the various processes in the two clusters. ICM related mechanisms,
such as ram-pressure stripping and starvation, have larger spheres of
influence in MS0451 than in CL0024.

With so many exciting projects on their way, and the warm hospitality
of Torino, I hope that the organizing committee will reconvene another
meeting on the subject in the near future.

\begin{acknowledgments}
It is a pleasure to acknowledge the numerous and valuable
contributions of my collaborators -- Oliver Czoske, Alan Dressler,
Patrick Hudelot, Jean-Paul Kneib, Phil Marshall, Sean Moran, Priya
Natarajan, Gus Oemler, Graham Smith, Ian Smail -- to the work
presented here. I am especially grateful to Richard Ellis for his
invaluable scientific contributions and much appreciated advice and
support. Finally, I would like to thank the organizers of this meeting
for the exciting scientific program, and for inviting me. Financial
support from IAU and from NASA through Hubble Fellowship grant
HF-01167.01 is gratefully acknowledged.
\end{acknowledgments}

\end{document}